\documentstyle[prd,aps,epsfig,preprint]{revtex}
\begin{document}
\draft
\preprint{                                                   BARI-TH/276-97}
\title{         Fourier analysis of real-time, high-statistics		\\ 
			solar neutrino observations			}
\author{         G.\ L.\ Fogli, E.\ Lisi, and D.\ Montanino		}
\address{     Dipartimento di Fisica and Sezione INFN di Bari,		\\
                  Via Amendola 173, I-70126 Bari, Italy			}
\maketitle
\begin{abstract}
Solar neutrino oscillations with wavelengths comparable to the Earth-Sun
distance provide a viable explanations of the long-standing solar neutrino 
deficit. They imply a time-dependent modulation of the solar neutrino flux 
due to the eccentricity of the Earth orbit. Motivated by this testable 
prediction, we propose a Fourier analysis of the signal observable in 
real-time, solar neutrino experiments. We give the general expressions of 
the Fourier coefficients and of their correlated uncertainties in the 
presence of background. The expressions assume a particularly compact form 
in the case of two-flavor neutrino oscillations in vacuum. We discuss the 
sensitivity to the lowest harmonics of the new-generation, high-statistics 
experiments SuperKamiokande, Sudbury Neutrino Observatory, and Borexino.
\end{abstract}


	The phenomenon of neutrino oscillations in vacuum \cite{Po67} 
represents a possible solution \cite{Gl87,Kr92,Ph92} to the long-standing 
solar neutrino problem \cite{Ba89}. Considering for simplicity only the 
first two neutrino families $\nu_e$ and $\nu_\mu$, the $\nu_e$ survival 
probability $P$ at a distance $L$ from the Sun is given by
\begin{equation}
	P(E)	=	1-\frac{\sin^2{2\theta}}{2} 
			\left(1-\cos\frac{\delta m^2 L}{2 E} \right)\ ,
\label{eq:P}
\end{equation}
where $E$ is the neutrino energy, $\delta m^2$ is the neutrino
squared mass difference, and $\theta$ is the vacuum mixing angle.

	A tentative evidence for the oscillating term in Eq.~(\ref{eq:P}) 
comes from the $E$-dependence of the solar neutrino deficit \cite{Kr97} as 
inferred from the four pioneering solar $\nu$ experiments \cite{Four}. The 
ellipticity of the Earth's orbit implies a further, striking signature of 
the oscillation phenomenon, namely, the $L$-dependence of the observed flux 
(in addition to the trivial $1/L^2$ geometric factor) \cite{Pome}. 
In particular, the survival probability $P$ in Eq.~(\ref{eq:P}) is modulated 
in time by the periodic variation of $L$ which, at first order in the 
eccentricity ($\varepsilon=0.0167$), is given by
\begin{equation}
	L(t)	=	L_0\,\left(1-\varepsilon \cos\frac{2\pi t}{T}\right)
				+{\cal O}(\varepsilon^2)
\label{eq:L}
\end{equation}
with $L_0=1$ AU, $T=1{\rm\ yr}$,  and $t=0$ at the perihelion. 
The  first-generation experiments \cite{Four} have not collected
enough statistics to test the $L$-dependence of the solar $\nu$ flux
\cite{Ca95}. New-generation experiments should instead be able to
probe the structure of the neutrino signal in the time domain \cite{Kr95}.

	In this work we propose to study the neutrino signal in the 
frequency domain through a Fourier analysis of the periodic variations 
associated to flavor oscillations in vacuum. This approach is particularly 
suited with real-time, high statistics experiments such as SuperKamiokande 
\cite{SKam} (operating), the Sudbury Neutrino Observatory \cite{SNOb}
(SNO, in construction), and Borexino \cite{Bore} (in construction). 
It will be seen that the different sensitivity of each experiment to the 
lowest harmonics  can help in discriminating the value of $\delta m^2$
(should the vacuum oscillation solution be confirmed).  Compact expressions 
for the Fourier coefficients and for their uncertainties will be given.

	In general,  the neutrino event rate $R(t)$ at the time $t$ is the 
sum of a signal $S(t)$ and of a (supposedly constant) background $B$,
\begin{equation}
	R(t)	=	B+S(t)\ .
\end{equation}
For symmetry reasons, the analysis can be restricted to the time interval  
$[0,\,T/2]$. It is understood that  events collected in subsequent half-years 
must be symmetrically folded in this interval. The data sample consists then 
of $N$ events collected at different times  $\{t_i\}_{1\leq i \leq N}$, with  
$t_i\in [0,\,T/2]$ and $N$ equal to the total sum of background and signal 
events, $N=N_S+N_B$. Notice that, in general, the background-signal separation 
can be performed on the average rates, but not on an event-by-event basis.

	The expansion of the signal in terms of Fourier components $f_n$ 
is defined as
\begin{equation}
	S(t)	=	S\left( 1+2\sum_{n=1}^{\infty}
			f_n\cos\frac{2\pi n t}{T}\right)\ ,
\end{equation}
where $S$ is the time-averaged signal
\begin{equation}
	S	=	\frac{2}{T}\int^{T/2}_0\! dt\,S(t)\ . 
\end{equation}
The $n$-th harmonic corresponds to a period of $1/n$~yr.
The explicit form  of $f_n$ reads
\begin{eqnarray}
	f_n 	& = & 	\frac{2}{ST} \int^{T/2}_0\! dt\,
			R(t) \cos\frac{2\pi n t}{T}
\label{eq:fntheo}\\
 		& = & 	\frac{1}{N_S}\sum_{i=1}^{N_S+N_B}
 			\cos\frac{2\pi n t_i}{T}\ ,
\label{eq:fnexpt}
\end{eqnarray}
where Eqs.~(\ref{eq:fntheo}) and (\ref{eq:fnexpt}) represent the theoretical
definition and the experimental determination of the $f_n$'s, respectively. 
Although in Eq.~(\ref{eq:fntheo}) one can replace $R(t)$ with $S(t)$,
the use of $R(t)$ is more general, since background events
do contribute to the statistical uncertainties. Notice that the sum in 
Eq.~(\ref{eq:fnexpt}) does not require an event-by-event separation of 
signal and background, and does not involve any binning in time.

	Each $f_n$ is a linear combination  of Poisson random variables 
$\xi_t=R(t)dt$ [Eq.~(\ref{eq:fntheo})], which represent the total number of 
events collected in the interval $[t,\,t+ dt]$. If only statistical errors 
are considered, then ${\rm var}(\xi_t)=\xi_t$ and 
${\rm cov}(\xi_t,\,\xi_{t'})=0$, since fluctuations at different times $t$
and $t'$ are uncorrelated \cite{Corr}. The  (co)variances of linear 
combinations of independent random variables  are given by \cite{Stat}
	${\rm var}(\sum_t \alpha_t\xi_t)=\sum_t \alpha_t^2{\rm var}(\xi_t)$ 
and
	${\rm cov}(\sum_t \alpha_t\xi_t,\,\sum_{t'}\beta_{t'}\xi_{t'})=
	\sum_t \alpha_t\beta_t{\rm var}(\xi_t)$. 
It follows that
$$
	{\rm var}(f_n)	=    \int_0^{T/2}\!\!dt\,R(t)\left(
			     \frac{2}{ST}\cos\frac{2\pi n t}{T}\right)^2\ ,
$$
$$
	{\rm cov}(f_n,\,f_m)=\int_0^{T/2}\!\!dt\,R(t)\,\frac{4}{(ST)^2}
			     \cos\frac{2\pi n t}{T}\cos\frac{2\pi m t}{T}\ .
$$
The final result for the one-sigma error $\sigma_n$ affecting $f_n$ and for 
the correlation $\rho_{mn}$ $(m\neq n)$ is:
\begin{equation}
	\sigma_n	=	\sqrt{\frac{1+f_{2n}+B/S}{ST}}\ ,
\label{eq:s}
\end{equation}
\begin{equation}
	\rho_{mn}	=	\frac{f_{m+n}+f_{|m-n|}}
				{\sqrt{(1+f_{2m}+B/S)(1+f_{2n}+B/S)}}\ .
\label{eq:rho}
\end{equation}
The values of $\sigma_n$ and $\rho_{mn}$ can be expressed in terms of measured
quantities with the substitutions $ST/2\to N_S$ and $B/S\to N_B/N_S$.

	In the standard (std) case, i.e.\ in the absence of oscillations, 
the Fourier transform of the signal
\begin{eqnarray}
	S(t)	\propto \left(\frac{L_0}{L(t)}\right)^2
		=	1+2\varepsilon\cos\frac{2\pi t}{T}
\end{eqnarray}
is trivial, only the first coefficient being nonzero and equal to the 
eccentricity $\varepsilon$,
\begin{equation}
	f_n^{\rm std} 	=  	\varepsilon\delta_{n1}\ .
\end{equation}
Moreover, in the standard case the statistical errors  do not depend on $n$,
\begin{equation}
	\sigma^{\rm std}_n  	= 	\sqrt{ \frac{N_B+N_S}{2N_S^2}}\ ,
\label{eq:sigma}
\end{equation}
and thus form a ``white noise'' affecting all the harmonics. The error
correlations, as derived from Eq.~(\ref{eq:rho}), are given by
$\rho^{\rm std}_{mn}=(\varepsilon N_S/N)\delta_{m,n\pm 1}\leq\varepsilon$ 
and thus are practically negligible.

	In the case of two-flavor oscillations the signal is proportional to
\begin{equation}
	S(t)	\propto \frac{L_0^2}{L^2(t)}\int\! dE\, \lambda(E) 
			\left[ \sigma_e(E) P + \sigma_\mu(E) (1-P)\right],
\label{eq:Sosc}
\end{equation}
where $\lambda(E)$ is the neutrino energy spectrum, $\sigma_e(E)$ and 
$\sigma_\mu(E)$ are the $\nu_e$ and $\nu_\mu$ interaction cross sections, 
and the probability $P$ is  given by Eq.~(\ref{eq:P}) with $L$ as in 
Eq.~(\ref{eq:L}). (It is understood that $\sigma_e$ and $\sigma_\mu$ are 
corrected for energy threshold and resolution effects in each detector.)

Given the signal $S(t)$ as in Eq.~(\ref{eq:Sosc}), the time integration
in the expression of the  Fourier coefficients
$$
	f_n	=	\frac{\int^{T/2}_0\!\!dt\,S(t)\,\cos
			\frac{2\pi n t}{T}}{\int^{T/2}_0\!\!dt\,S(t)}\ 
$$
can be performed analytically. The final result can be cast
in the following, compact form:
\begin{equation}
	f_n	=	\frac{\varepsilon\delta_{n1}-\sin^2{2\theta}
			\,D_n(\delta m^2)}{1-\sin^2{2\theta}\,D_0
			(\delta m^2)}\;\;\;\;\;(n\geq 1)\ ,
\label{eq:fnosc}
\end{equation}
where the detector-dependent functions $D_n$ are given by
\begin{equation}
	D_n(\delta m^2)=\frac{\int\!dE\,\lambda\,(\sigma_e-\sigma_\mu)\,U_n}
	{2\int\!dE\,\lambda\,\sigma_e}\;\;\;(n\geq 0)
\end{equation}
and the universal (i.e., detector-independent) functions $U_n$ are
given by
$$
	U_n(z)	=	\delta_{n0}-u_n(z)-\varepsilon
			[u_{n+1}(z)+u_{n-1}(z)-\delta_{n1}]\ ,
$$
$$
	u_n(z)	=	\cos\left(z-\frac{n\pi}{2}\right)J_n(\varepsilon z)\ ,
$$
where $z=\delta m^2 L_0/2 E$ and $J_n$ is the Bessel function of order $n$ 
\cite{Math}. Notice that, although our calculations are of 
${\cal O}( \varepsilon$), all orders in $\varepsilon z$ are kept, since 
$z$ may be large.

	We apply now the Fourier analysis to the signal expected in the
SuperKamiokande, SNO, and Borexino experiments. The SuperKamiokande and 
Borexino cross sections $(\nu +e\to \nu' +e)$ are taken from \cite{SKcr}, 
and the SNO cross section $(\nu_e+ d \to p +p+e)$ from \cite{SNcr}. These 
cross sections are corrected for energy threshold and resolution effects
(see, e.g., \cite{Ba97}). In particular, we consider a prospected threshold 
of 5 MeV for the recoil  electron kinetic energy $T_e$ in SuperKamiokande and 
SNO, and an analysis  window $T_e\in[0.25,\,0.8]$ MeV for Borexino. The energy 
resolution is assumed to be $16\%$ at 10 MeV for SuperKamiokande, $11\%$ at 
10 MeV for SNO, and $8\%$ at 0.5 MeV for Borexino, scaling as $\sqrt{T_e}$ for 
different recoil energies. The Fourier components of the vacuum oscillation 
signal are calculated through Eq.~(\ref{eq:fnosc}) in the $\delta m^2$ range  
relevant for the solution of the solar neutrino problem and then plotted 
in Fig.~1.

	Figure~1 shows the deviations of the Fourier components from their 
standard values, $f_n-\varepsilon\delta_{n1}$, as functions of $\delta m^2$
for maximal mixing ($\sin^2{2\theta}=1$).   The deviations decrease  with 
decreasing mixing (not shown), although not exactly in  proportion to 
$\sin^2{2\theta}$ [see Eq.~(\ref{eq:fnosc})]. The three subfigures refer 
to SuperKamiokande, SNO, and Borexino (top to bottom). The gray, horizontal 
bands correspond  to the $\pm 1\sigma$ statistical error  relative to the 
standard (no oscillation) case for $N_S=10^4$ events and 
$N_B=0$ $(\sigma_n= 0.0071)$.  Error bands for different choices of 
$N_S$ or $N_B$ can be obtained through Eq.~(\ref{eq:sigma}).

	It can be seen from Fig.~1 that the SuperKamiokande and SNO experiments 
are sensitive  only to the first harmonic, i.e. to variations of the signal 
with a period of 1 yr. The amplitude of the second harmonic is, in fact, 
much smaller   than the $\pm 1\sigma$ error in both cases. 
This is in agreement with the results in \cite{Kr95}, where the SNO and 
SuperKamiokande signals were shown to vary almost sinusoidally in time. 
A single parameter $(f_1)$ is then sufficient to characterize the results 
of each of these two experiments. For $10^4$  events and no background, 
Fig.~1 shows that the genuine $n=1$ oscillation component $(f_1-\varepsilon)$
may reach a significance of $\sim3\sigma$ ($\sim7\sigma$) in the most 
favorable case for SuperKamiokande (SNO). For a signal-to-background
ratio equal to one ($N_B=N_S$), the statistical significance of the vacuum 
oscillation signal would be lower by a factor $\sqrt{2}$. This implies that 
the detection of an unmistakable signal for vacuum oscillation in the time or 
frequency domain  will require very high statistics and  good background 
rejection in both experiments. Even in the best conditions, there are some 
ranges of $\delta m^2$  where the expected time-modulation 
of the signal is small and undetectable (e.g., for  
$\delta m^2 \simeq 1.6\times 10^{-10}{\rm\ eV}^2$ in Fig.~1) \cite{Kr95,Comm}.
In such ranges, however, independent oscillation signals might  show up
in the energy domain as distortions of the recoil electron spectrum 
\cite{Kr95,Ba97}.

	The situation is much more favorable for the Borexino experiment,
since its signal is dominated by the monoenergetic $^7$Be solar neutrinos
($E=0.862$ MeV), while the SuperKamiokande and SNO signals are smeared by 
the broad $^8$B  neutrino spectrum ($E\lesssim 15$ MeV).  The third panel 
of Fig.~1 shows, in fact, that at least one of the first three Fourier 
coefficients is larger than the (representative) statistical error in the 
$\delta m^2$ range of interest. Moreover, when one of the harmonics is small, 
another is large (and vice versa), so that no ``holes'' are left
in the sensitivity to the $\delta m^2$ variable, provided that (at least) 
two of the first three Fourier components are measured. Notice also that
the relative amplitude of the Fourier coefficients in Borexino is strongly
dependent on $\delta m^2$, and thus the detection of two nonzero harmonics 
would be of great help in discriminating a preferred range of $\delta m^2$. 
The relative amplitude of the first Fourier coefficient in SuperKamiokande 
(or SNO) and Borexino is also dependent on $\delta m^2$ and, therefore, the 
combination of all the experiments will enhance the resolution in 
$\delta m^2$.

	In conclusion, we have performed a Fourier analysis of the
signal expected in the SuperKamiokande, SNO, and Borexino solar neutrino 
experiments. This method fully exploits the real-time features of the three 
detectors and requires no binning in time. Expressions for the correlated
uncertainties of the Fourier components have been worked out in the general 
case and, in particular, for no oscillation. Compact expressions for the 
Fourier coefficients in the presence of $2\nu$ oscillations have been given.
The method has been applied to the analysis of the signals
expected in the new-generation experiments. SuperKamiokande and SNO are 
shown to be sensitive  only to the first harmonic, while Borexino is also 
sensitive to the second and third harmonic. The combination of 
different Fourier component measurements is highly selective in $\delta m^2$.

	We thank G.\ Bellini, M.\ G.\ Giammarchi,  and A.\ Ianni
for useful correspondence, and P.\ I.\ Krastev for helpful suggestions.



\begin{figure}
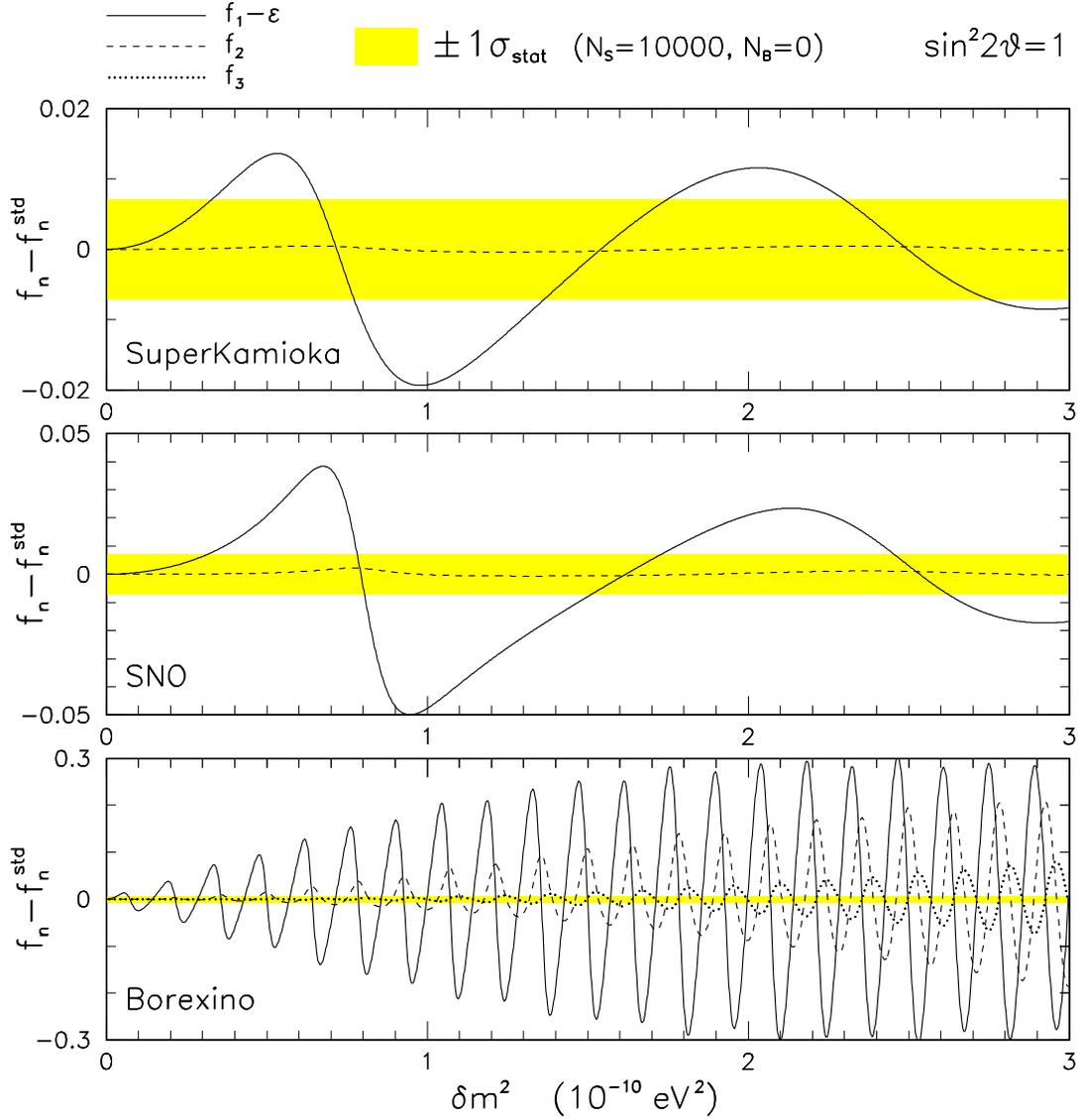

\caption{Deviations of the Fourier coefficients $f_n$ from their standard 
	values $f_n^{\rm std}=\varepsilon\delta_{n1}$, as functions of 
	$\delta m^2$  for maximal $2\nu$ mixing ($\sin^2{2\theta}=1$). 
	In each of the three panels (SuperKamiokande, SNO, and Borexino), 
	the gray, horizontal band represents the $\pm 1\sigma$ statistical 
	uncertainty associated to the standard predictions $f_n^{\rm std}$
	for $10^4$ events with no background.  SuperKamiokande and SNO
	appear to be sensitive only to the first harmonic (solid curve), 
	while Borexino is sensitive also to the second and third harmonic 
	(dashed and dotted curve, respectively). Notice that the vertical 
	scales are different, but the absolute width of the gray error 
	band $(\sigma_n=\pm0.0071)$ is the same for the three experiments.}
\end{figure}



\newcommand{\InsertFigure}[2]{\newpage\begin{center}\mbox{%
\epsfig{bbllx=1.4truecm,bblly=1.3truecm,bburx=19.5truecm,bbury=26.5truecm,%
height=21.truecm,figure=#1}}\end{center}\vspace*{-2.85truecm}%
\parbox[t]{\hsize}{\small\baselineskip=0.5truecm\hskip0.5truecm #2}}
\InsertFigure{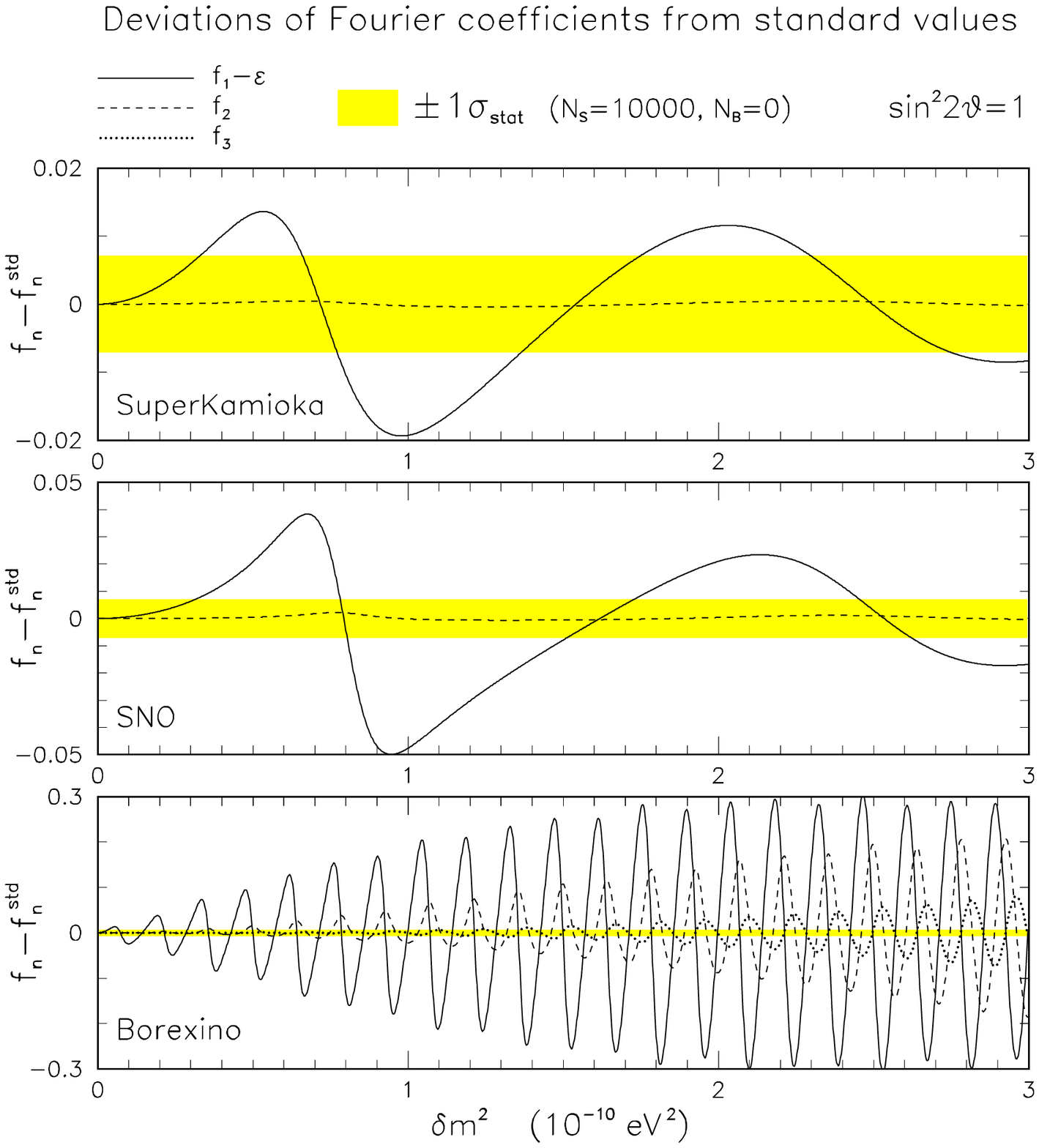}%
{FIG.~1  Deviations of the Fourier coefficients $f_n$ from their standard 
	values $f_n^{\rm std}=\varepsilon\delta_{n1}$, as functions of 
	$\delta m^2$  for maximal $2\nu$ mixing ($\sin^2{2\theta}=1$). 
	In each of the three panels (SuperKamiokande, SNO, and Borexino), 
	the gray, horizontal band represents the $\pm 1\sigma$ statistical 
	uncertainty associated to the standard predictions $f_n^{\rm std}$
	for $10^4$ events with no background.  SuperKamiokande and SNO
	appear to be sensitive only to the first harmonic (solid curve), 
	while Borexino is sensitive also to the second and third harmonic 
	(dashed and dotted curve, respectively). Notice that the vertical 
	scales are different, but the absolute width of the gray error 
	band $(\sigma_n=\pm0.0071)$ is the same for the three experiments.}

\begin{thebibliography}{99}
\bibitem{Corr}	We do not consider additional systematic effects that might 
		be correlated in time, e.g., possible cyclic variations of 
		the detector efficiency. The evaluation of these experimental
		uncertainties is premature and goes beyond the scope of this 
		work.
\end{thebibliography}
\end{document}